\documentclass[twocolumn]{aastex63}
\usepackage{amssymb}
\usepackage{array}
\usepackage{multirow}
\usepackage{bold-extra}
\AtBeginDocument{
}

\received{2020 September 18}
\revised{2020 November 17}
\accepted{2020 November 26}

\submitjournal{ApJ}

\shorttitle{TeV to PeV neutrinos from radio blazars}
\shortauthors{Plavin et al.}
\graphicspath{{./}{figures/}}

\begin{document}

\title{Directional association of TeV to PeV astrophysical neutrinos with radio blazars}

\correspondingauthor{Alexander Plavin}
\email{alexander@plav.in}

\author[0000-0003-2914-8554]{A.~V.~Plavin}
\affiliation{Astro Space Center of Lebedev Physical Institute, Profsoyuznaya 84/32, 117997 Moscow, Russia}
\affiliation{Moscow Institute of Physics and Technology, Institutsky per. 9, Dolgoprudny 141700, Russia}

\author[0000-0001-9303-3263]{Y.~Y.~Kovalev}
\affiliation{Astro Space Center of Lebedev Physical Institute, Profsoyuznaya 84/32, 117997 Moscow, Russia}
\affiliation{Moscow Institute of Physics and Technology, Institutsky per. 9, Dolgoprudny 141700, Russia}
\affiliation{Max-Planck-Institut f\"ur Radioastronomie, Auf dem H\"ugel 69, 53121 Bonn, Germany}

\author[0000-0002-8017-5665]{Yu.~A.~Kovalev}
\affiliation{Astro Space Center of Lebedev Physical Institute, Profsoyuznaya 84/32, 117997 Moscow, Russia}

\author[0000-0001-6917-6600]{S.~V.~Troitsky}
\affiliation{Institute for Nuclear Research of the Russian Academy of Sciences, 60th October Anniversary Prospect 7a, Moscow 117312, Russia
}

\begin{abstract}
Recently we have shown that high-energy neutrinos above 200~TeV detected by IceCube are produced within several parsecs in the central regions of radio-bright blazars, that is active galactic nuclei with jets pointing towards us. To independently test this result and extend the analysis to a wider energy range, we use public data for all neutrino energies from seven years of IceCube observations. The IceCube point-source likelihood map is analyzed against the positions of blazars from a statistically complete sample selected by their compact radio flux density. The latter analysis delivers a $3.0\sigma$ significance with the combined post-trial significance of both studies being $4.1\sigma$. 
The correlation is driven by a large number of blazars. Together with fainter but physically similar sources not included in the sample, they may explain the entire IceCube astrophysical neutrino flux as derived from muon-track analyses.
The neutrinos can be produced in interactions of relativistic protons with X-ray self-Compton photons in parsec-scale blazar jets.
\end{abstract}


\section{Introduction} \label{s:intro}

The origin of high-energy astrophysical neutrinos, being detected for years by the IceCube experiment  \citep[e.g.,][]{AhlersHalzen} was puzzling. The observational data do not support some of the earlier expectations and put forward more puzzles, thus, complicating theoretical implications. Active galactic nuclei (AGNs) and blazars in particular have been considered as a probable class of the neutrino sources since the very early days of the multimessenger astronomy \citep{BerezinskyNeutrino77,Eichler1979,BerezGinzb}. However, no statistically significant association of neutrino events with gamma-ray loud blazars has been found \citep[see, e.g.,][]{corr-1611.03874IceCube, corr-1611.06338Neronov, corr-1702.08779Vissani, corr-1807.04299Tavecchio, corr-1908.08458IceCube,aartsenTimeIntegratedNeutrinoSource2020}. In addition, gamma-ray blazars are not numerous, and the lack of observation of individual bright sources puts strong constraints on this scenario, cf.\ \citet{MuraseCombined,NeronovEvolution,Finley-2005.02395}. 
This contrasts with the association of a single high-energy neutrino event \citep{IceCubeTXSgamma} with a gamma-ray flare in the blazar TXS~0506+056 and an excess of low-energy events from the same direction \citep{icecubecollaborationNeutrinoEmissionDirection2018}. Detailed discussions of the AGN models of astrophysical neutrinos can be found, e.g., in reviews by \citet{Boettcher-rev,Cerruti-rev}, while more general descriptions of IceCube observations and of various scenarios of the neutrino origin are presented, e.g., by \citet{AhlersHalzen,VissaniUniverse}. In this work we use the term ``blazar'' to denote an AGN with the jet pointed towards the observer at a viewing angle of several degrees following the unified scheme in \citet{1995PASP..107..803U}. This naming does not directly imply any specific spectral or other properties.

In \citet{neutradio1}, we recently have found a $3.1\sigma$-significant association of track-like events with estimated neutrino energies $E_\nu >200$~TeV from publicly available IceCube lists with radio-bright compact parsec-scale blazar cores. The crucial point in this observation was the use of very-long-baseline radio interferometry (VLBI) capable of resolving AGNs' central parsecs. Particularly, we utilized a complete full-sky flux-density-limited sample of VLBI-selected extragalactic sources. We found that blazars whose positions in the sky coincide with arrival directions of neutrino events have stronger parsec-scale radio cores than the rest of the sample, with the post-trial probability of random coincidence of $2\cdot10^{-3}$. Moreover, on average, these potential neutrino sources exhibited radio flares at the time of the neutrino arrival as measured by the RATAN-600 radio telescope \citep{neutradio1} and confirmed later with Owens Valley and Metsähovi observations by \citet{hovatta2020}. Besides these statistical analyses, a possible temporal correlation of neutrino events and radio flares was pointed out for two particular individual sources by \citet{Kadler16,IceCubeTXSgamma,kun2020}.

Those observations reopened the possibility of explaining the bulk of high-energy astrophysical neutrinos by blazars while evading constraints from the lack of both gamma-ray associations and significant individual point-like neutrino sources. In addition, this result suggests physical grounds for the association of TXS~0506$+$056, a typical blazar bright both in radio \citep{r:kovalev0506,r:ros0506} and in gamma rays, with the IceCube 170922A neutrino event \citep{IceCubeTXSgamma}. This object looks orphan otherwise in the absence of other similar neutrino/gamma-ray coincidences.

After the analysis results were published by \citet{neutradio1}, new observational data became available. The IceCube Collaboration has released seven-year (2008--2015) public data adopted for the search of point-like neutrino sources \citep{icecubecollaborationAllskyPointsourceIceCube}. The main task of our present work is to utilize these newly released neutrino data to search for associations with VLBI-selected blazars from the sample used by \citet{neutradio1}. Besides the possibility to independently test the conjectured neutrino~-- radio blazars association, this analysis is potentially important for astrophysical conclusions. The new data we use here are based on neutrino detections in a wide energy range starting from TeVs. In contrast, the events above 200~TeV, used in the previous analysis, represent only the very tail of the IceCube spectrum, cf.\ \citet{1908.09551,2020arXiv200109520I}. The production of TeV and sub-PeV neutrinos in AGNs may require different physical conditions. Hence the present study provides additional nontrivial constraints on theoretical models. 

The rest of the paper is organized as follows. In \autoref{s:data}, we describe the public IceCube (\autoref{s:data_icecube}) and VLBI (\autoref{s:data_vlbi}) data used in the study. \autoref{s:analysis_logp} presents the analysis of these data and its results. In \autoref{s:joint_analysis}, we compare these results with those of \citet{neutradio1}
and derive a combined statistical significance of the two studies.
\autoref{s:flux} estimates the neutrino flux produced by blazars and puts it into a wider context.
\autoref{s:discussion} discusses theoretical implications of our results. In \autoref{s:summary}, we briefly summarize our findings and discuss ongoing and future observations aimed to explore and refine the results of this study.
 
\section{Description of the utilized data} \label{s:data}

\begin{figure*}
    \centering
    \includegraphics[width=\linewidth]{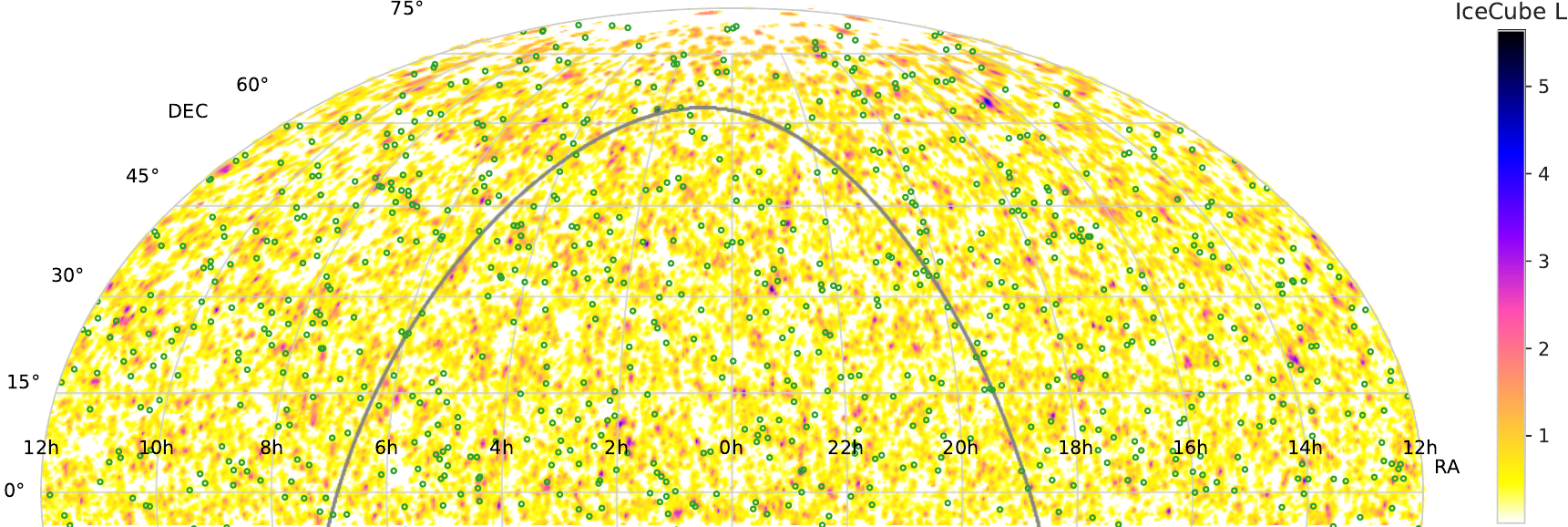}
    \caption{Sky map in equatorial coordinates of the IceCube local $p$-value logarithms denoted as $L$. Darker areas with larger $L$ indicate higher probabilities to have an astrophysical neutrino point source in this direction, see \autoref{s:data_icecube}. All sky north of $\delta = -5^\circ$ is displayed in equatorial coordinates. Radio AGNs from the complete 8~GHz VLBI sample down to the flux density of 0.33~Jy are shown as green circles. The grey line represents the Galactic plane.}
    \label{fig:skymap}
\end{figure*}

\subsection{IceCube Neutrino Detections} \label{s:data_icecube}

The IceCube Collaboration routinely performs dedicated searches for point-like neutrino sources as the data are accumulated \citep{icecubecollaborationAllskySearchTimeintegrated2017,2019EPJC...79..234A,2020APh...11602392A,IceCube-1907.06714,aartsenTimeIntegratedNeutrinoSource2020}, also jointly with ANTARES \citep{IceCube-ANTARES-2001.04412}. No significant individual point source was found: in all cases, the post-trial significance is below $3\sigma$. The most significant source from a predefined catalog is a star-forming Seyfert galaxy NGC~1068 \citep[$2.9\sigma$ post-trial with 10 years of track-like data,][]{aartsenTimeIntegratedNeutrinoSource2020}.  Stacking analyses of radio-selected sources were not performed in those works.

There are two kinds of high-energy neutrino detections at IceCube: cascades and tracks. Cascades arise from showers developing within the volume of the detector itself, and tracks are registered when secondary muons pass through the detector. The full ten-year IceCube track data, as well as cascade data, have not been publicly released yet. However, the track data covering seven years (2008-2015) used in the analysis of \citet{icecubecollaborationAllskySearchTimeintegrated2017} has been published \citep{icecubecollaborationAllskyPointsourceIceCube} in the form of pre-trial local $p$-values on a grid of pixels covering the entire sky. 
We denote the provided negative logarithms of local $p$-values as $L = -\log p$, so that they are not confused with statistical $p$-values of our present analysis computed in the following sections of this work. Our analysis treats $L$ as a measure of the detected direction-dependent neutrino emission: a larger $L$ corresponds to a higher probability that a point source of astrophysical neutrinos is located in a given direction in the sky.

The $L$ values map is based on 712830 detected events but does not explicitly contain the information on their individual properties. These $L$ are based on the likelihood of the model assuming an astrophysical neutrino source in a given direction with a power-law spectrum. Therefore, they accumulate information about arrival directions and energies of neutrinos in an area of the sky over the entire observational period of seven years. Note that the probability that an event has an astrophysical origin and is not caused by the atmospheric background grows with energy. This dependency was included in the calculation of the likelihood \citep{icecubecollaborationAllskySearchTimeintegrated2017}. The neutrino energies further influence the likelihood through the energy-dependent angular resolution and sensitivity of IceCube. The effective area, reported in \citet{icecubecollaborationAllskyPointsourceIceCube}, is generally larger for more energetic particles, and steeply falls below $\sim 10$~TeV. The angular resolution is roughly $0.5^\circ$ at energies above 100~TeV, and increases for less energetic neutrinos. 
The distance between neighbouring pixels in the grid is about $0.1^\circ$, several times smaller than the highest achieved resolution; therefore, we do not perform any additional oversampling or interpolation. We exclude the Southern sky (declination $\delta < -5^\circ$) from our analysis. The sensitivity of IceCube to astrophysical neutrinos in this range, above the horizon when observed from the South Pole, is heavily degraded; see discussion of the effect in \citep{icecubecollaborationAllskySearchTimeintegrated2017}. Further in this work, we refer to the $\delta>-5^\circ$ range as the Northern sky. This area of the map is influenced by 422791 individual detection events \citep{icecubecollaborationAllskySearchTimeintegrated2017}.

In our work, we additionally utilize the largest published dataset of individual IceCube events that covers only three years, 2010-2012 \citep{icecubecollaborationAllskyPointsourceIceCube2018}. It contains 334677 events, and 196316 of those are in the Northern sky. Each event in the catalog is described by the arrival direction in the sky, statistical uncertainties of this direction, and the estimated particle energy. We remove detections whose 90\% containment area on the celestial sphere is larger than $10$~deg$^2$, which leaves us with 114799 events north of $\delta = -5^\circ$ out of the original 196316.

\subsection{VLBI Observations of Blazars} \label{s:data_vlbi}

For our analysis we use the same 8~GHz VLBI data as in \citet{neutradio1}, which are compiled in the Astrogeo\footnote{\url{ http://astrogeo.org/vlbi_images/}} database. These observations include geodetic VLBI programs \citep{2009JGeod..83..859P,2012A&A...544A..34P,2012ApJ...758...84P}, the Very Long Baseline Array (VLBA) calibrator surveys (VCS; \citealt{2002ApJS..141...13B,2003AJ....126.2562F,2005AJ....129.1163P,2006AJ....131.1872P,2007AJ....133.1236K,2008AJ....136..580P,VCS9,2016AJ....151..154G}), and other 8~GHz global VLBI, VLBA, EVN (the European VLBI Network), LBA (the Australian Long Baseline Array) observations \citep{2011AJ....142...35P,2011AJ....142..105P,2011MNRAS.414.2528P,2012MNRAS.419.1097P,2013AJ....146....5P,2015ApJS..217....4S,2017ApJS..230...13S,2019MNRAS.485...88P,2020arXiv200809243P,2020arXiv200806803P}. The AGN positions and radio flux densities are determined from these observations and presented within the VLBI-based Radio Fundamental Catalogue\footnote{\url{http://astrogeo.org/rfc/}} (RFC). We use the latest available version to date, RFC~2020b.
The catalog includes a complete sample of AGNs limited by the 8~GHz flux density integrated over VLBI images $S_\mathrm{8GHz} \geq 150$~mJy. This complete sample contains 3411 objects, and 1938 of those are located north of $\delta = -5^\circ$.
For the sources with multiple VLBI observations, the individual flux density measurements are averaged. The median number of 8~GHz observations for a source is five, the maximal is more than 150.

In our analysis, we utilize the aforementioned flux densities of AGNs, call them ``VLBI flux densities'' throughout the paper, and denote as $S$ in equations. The sample is dominated by AGNs with strongly Doppler-boosted relativistic jets pointing towards the observer, and their VLBI flux density primarily consists of the parsec-scale jet emission \citep[e.g.,][]{kovalevSubMilliarcsecondImagingQuasars2005,2012A&A...544A..34P,JetAngles,2019ApJ...874...43L}. We reiterate that the name ``blazars'' is used for these beamed objects throughout this paper following the unified scheme for AGNs \citep{1995PASP..107..803U}.

\section{Directional Correlation of radio blazars with time-integrated neutrino flux} \label{s:analysis_logp}

Here, we address the question of whether radio-bright VLBI-selected blazars are associated with neutrinos detected by IceCube. We utilize 
the time-aggregated likelihoods of astrophysical neutrino sources from IceCube (\autoref{s:data_icecube}) together with the average historic VLBI flux density of blazars (\autoref{s:data_vlbi}). \autoref{app:icecube_map} motivates and discusses the $L$-value normalization that results in $L_{\rm norm}$ values used in our analysis.

\begin{figure}
    \centering
    \includegraphics[width=\linewidth]{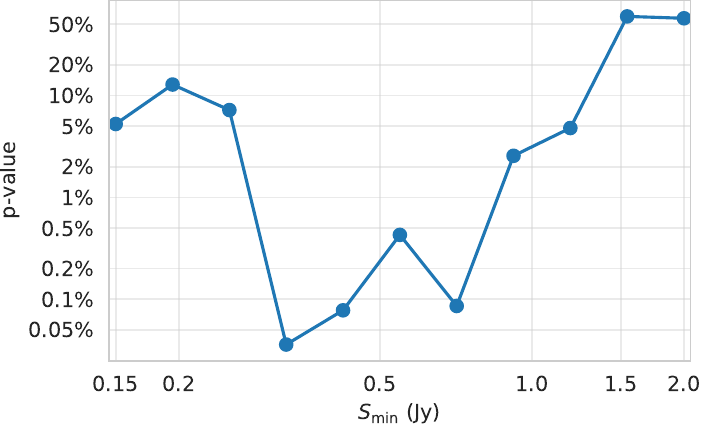}
    \caption{$P$-values for a range of VLBI flux density cutoffs. Each value indicates the probability that the observed correlation between neutrinos and blazars stronger than the flux cutoff happened by a random chance; see \autoref{s:analysis_logp}. The cutoffs values $S_{\rm min}$ split the interval 0.15-2 Jy into ten parts uniformly in log-scale. The lowest $p$-value of $4\cdot10^{-4}$ is attained for $S_{\rm min} = 0.33$~Jy.}
    \label{fig:logp_pval_flux}
\end{figure}

We select all blazars with the VLBI flux density higher than a threshold $S_{\rm min}$ determined below, extract $L_{\rm norm}$ values at their positions, and take the median of these values as the test statistic. Then we test whether this statistic's value is higher than could arise by chance: the same calculations are performed for all blazar positions randomly shifted in Right Ascension. This procedure is repeated $10^5$~times, yielding the null distribution. The pre-trial $p$-values are calculated based on this. See our previous study \citep{neutradio1} for motivation and details of this testing procedure. The trial $S_{\rm min}$ thresholds are taken from the interval [150~mJy; 2~Jy] split into ten parts evenly in logarithmic scale. Here, the lower bound is dictated by the sample completeness (\autoref{s:data_vlbi}), and the upper bound is chosen so that at least 40 objects remain in the selection. The $p$-values for each threshold are shown in \autoref{fig:logp_pval_flux}: the minimum of $p=4\cdot10^{-4}$ is achieved for $S_{\rm min} = 0.33$~Jy. Then the global post-trial $p$-value is calculated following the motivation and procedure detailed in \citet{neutradio1}.
It represents the chance probability for the median $L_{\rm norm}$ at bright blazar positions to be as high as actually observed. This probability is found to be $3\cdot10^{-3}$, which corresponds to the significance of $3.0\sigma$ for a normal distribution. 

The particular value of $S_{\rm min}$ yielding the minimal $p$-value is unlikely to bear any specific astrophysical meaning: it represents a trade-off between fewer sources remaining with higher cutoffs, and fainter sources covering larger fractions of the sky at lower thresholds. This trade-off depends on the sensitivity and effective resolution of neutrino telescopes, and the optimal $S_{\rm min}$ will likely be different for different datasets. However, for the available observational data from IceCube, we find that bright blazars above 0.33~Jy dominate the association with neutrinos. This value is used in \autoref{s:flux:energy} for rough estimates of the neutrino flux from blazars.

\begin{figure}
    \centering
    \includegraphics[width=\linewidth]{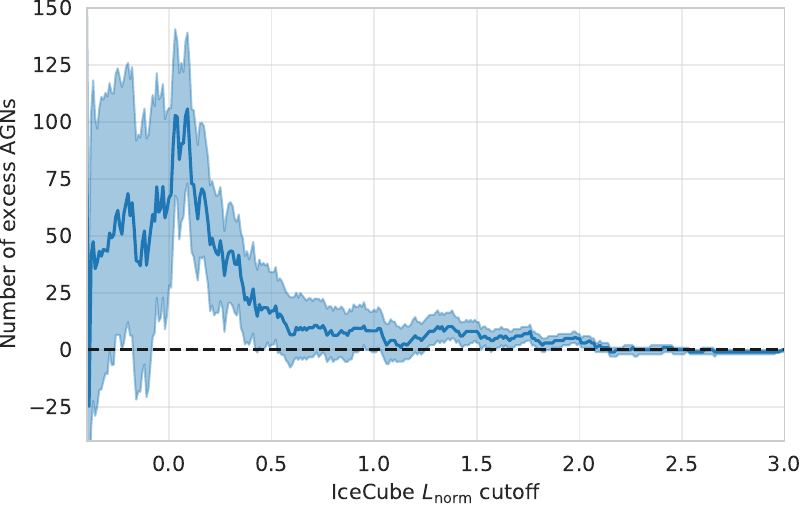}
    \caption{Number of excess blazars with normalized IceCube $L_{\rm norm}$ values higher than a given cutoff. The blue line indicates the estimated number itself, and the shaded area represents the $1\sigma$ uncertainty. These counts are estimates of how many blazars are associated with detected neutrinos, if only objects with $L_{\rm norm}$ higher than the cutoff are considered.}
    \label{fig:logp_agncnt}
\end{figure}

After having determined with the significance of $3.0\sigma$ that an excess of neutrinos is detected from the directions of radio-bright blazars, we estimate how many extragalactic objects drive this correlation. For a range of $L_{\rm norm}$ thresholds, we count the number of blazars with $L_{\rm norm}$ at their position in the sky higher than the threshold. Then we subtract the counts obtained in the same way for blazar positions randomly shifted in Right Ascension. These differences yield estimates of the excess number of blazars that are shown in \autoref{fig:logp_agncnt}. The maximum excess count of $104 \pm 32$ objects is achieved for $L_{\rm norm} > 0.09$: a value close to zero, and zero is the median of $L_{\rm norm}$. Together with the monotone decrease of the excess number for higher $L_{\rm norm}$ thresholds, this suggests that the majority of neutrino-emitting blazars have IceCube $L$ values that are not extremely high. Instead, their $L$ values are close to the overall median. Such blazars can only be distinguished by a statistical approach of this kind, and would be lost in any analysis focused on the brightest regions of the IceCube map.

This maximum excess count can slightly overestimate the true number of responsible blazars because of fluctuations: this reflects the multiple comparisons problem. Thus we proceed with a conservative approach. Maxima of the excess are computed for blazar positions randomly shifted in Right Ascension, following the same motivation as above. These maxima are subtracted from the real-positions maximum. The resulting estimate of the detected excess count is $72 \pm 23$ objects. We take this as an estimate of the number of blazars that significantly contribute to the correlation with neutrinos established above. Note that there are likely more neutrino-emitting blazars, but their neutrino flux is too low to be detected even in this aggregated fashion. Thus the excess counts are expected to go up when more sensitive observations become available. The current values are to be treated as lower limits.

\section{Selected high energy events and the all-sky IceCube map: joint analysis}
\label{s:joint_analysis}

Earlier this year in \citet{neutradio1}, we showed that blazars positionally associated with IceCube events above 200~TeV have brighter parsec-scale cores than the rest of the blazar sample. The analysis in the present work is based on IceCube data released after that paper was published, and thus can be considered a largely independent statistical study. The effects explored in this and the previous paper correspond to the same kind of objects, same spatial scales, and underlying mechanisms; the main difference is the range of neutrino energies dictated by the available observational data from IceCube.

After the work of \citet{neutradio1} was published, it has been pointed out \citep{hovatta2020} that the IceCube event on 2012-05-15 has the reported energy of exactly 200~TeV. This event was not used in our original analysis due to the strict inequality requirement $E>200$~TeV. We consider this being non-optimal and repeat the analysis of \citet{neutradio1}, including this event. The arrival direction of this neutrino is positionally associated with a compact radio-bright quasar 1308+326 (OP~313) that has the average VLBI flux density of 1.9~Jy at 8~GHz. Additionally, this blazar experienced a major radio flare peaking around 2012, temporally coincident with the neutrino detection \citep{hovatta2020}. Including this 200~TeV event slightly increases the post-trial statistical significance obtained in the analysis in \cite{neutradio1} from $3.1\sigma$ ($p=2\cdot10^{-3}$) to $3.4\sigma$ ($p=7\cdot10^{-4}$). We use this updated $p$-value below to improve the  completeness of the neutrino-VLBI analysis covering high energy events in 2010-2019. This choice does not affect any of our conclusions.

Further in this section we combine the results of \citet{neutradio1} and of \autoref{s:analysis_logp} to obtain the joint significance level of the neutrino-blazar association. This combination is justified because each of the two utilized IceCube datasets contains important information not available in the other. The catalog used in \cite{neutradio1} covers years from 2010 to 2019 and contains detailed information about each event such as its direction and energy. However, it lists only high-energy neutrinos that have a high probability to be of astrophysical origin. The analysis in \autoref{s:analysis_logp}, conversely, is based on all detected events independently of their energy, but only contains highly aggregated information and covers the period from 2008 to 2015. To perform the combination in the most conservative way possible, we apply a mask to the IceCube map of $L$ (see \autoref{s:data_icecube} for the map details) to remove the contribution of individual neutrinos already accounted in the previous analysis. Specifically, we mask out the pixels that are close to any of the IceCube events earlier than 2016.0 listed in \cite{neutradio1} with an addition of the 2012-05-15 event discussed above. The closeness threshold is taken equal to the positional errors of those events assuming a $0.5^\circ$ systematic uncertainty, see \cite{neutradio1} for discussion. With these regions masked out, the post-trial $p$-value based on the all-sky map analysis described in \autoref{s:analysis_logp} marginally increases from $3\cdot10^{-3}$ to $4\cdot10^{-3}$. Finally, we follow the Fischer's method of combining $p$-values from independent analyses \citep{fisher} and obtain the joint probability of chance coincidence $p=4\cdot10^{-5}$ that corresponds to $4.1\sigma$ for a normal distribution.

The physical effects and source regions probed by our analyses of higher- and lower-energy neutrino samples are the same. At the same time, the number of blazars manifesting themselves as neutrino emitters differs significantly. Neutrinos above 200~TeV are naturally rare: there are only 57 events passing the selection criteria in \citet{neutradio1}. About 35 of them are expected to have an astrophysical origin, and about 10 blazars from our sample can be currently associated with these events \citep{neutradio1}. Those objects are among the brightest or the most flaring radio blazars in the sky. Lower energy neutrinos are much more numerous with 422791 events in the Northern sky contributing to the IceCube map (\autoref{s:data_icecube}, \autoref{fig:skymap}). Only a fraction of $3 \cdot 10^{-3}$ of them, which is around 1300, are likely to have an astrophysical origin, cf.\ \citet{1908.09551}. As illustrated in \autoref{fig:logp_agncnt} and deduced in \autoref{s:analysis_logp}, about 70~blazars can already be associated with such neutrinos, even though we cannot reliably list individual sources yet. They constitute a noticeable fraction of the entire compact radio-bright blazars population: there are 1938 objects in total within our complete sample in the Northern sky, and 725 of them are brighter than $0.33$~Jy, the threshold yielding the minimal $p$-value in \autoref{fig:logp_pval_flux}.

\section{Estimate of the total neutrino flux from blazars}
\label{s:flux}

\subsection{Counting Individual Neutrinos in the Three-Year Dataset}
\label{s:flux:count}

The number of neutrino-associated blazars alone does not allow us to estimate the neutrino flux produced by them: this requires counting individual detection events instead of objects. The all-sky map of IceCube $L$ does not contain enough information to do that. Thus, we attempt to derive these counts from a smaller three-year dataset of individual events detected at IceCube \citep[see \autoref{s:data_icecube} and][]{icecubecollaborationAllskyPointsourceIceCube2018}. Due to a shorter time period of three years compared to seven for the $L$ map, any estimates based on these data are expected to be noisier.

We count the number of events with at least one blazar from our complete sample above 0.15~Jy falling within positional uncertainties. This counting is repeated for the real dataset and randomly-shifted samples, utilizing the same approach as in \autoref{s:analysis_logp}. The difference of these counts yields the number of excess neutrinos, i.e., those that are associated with AGNs. This number is $215\pm134$ objects, which is consistent with zero at a $2\sigma$ level. Thus, it does not constitute a significant evidence of a neutrino-blazar connection by itself. The lower significance in this case could be expected: the dataset is based on fewer observations, and there is more noise at low neutrino energies resulting from the atmospheric background. This noise is suppressed when the map of $L$ values is computed by IceCube (\autoref{s:data_icecube}). Nevertheless, it is still instructive to compare this estimate of the number of blazar-associated neutrinos over the three years to the total number of detection events in the Northern sky over this period, 196316. We find that a fraction of $(1.1 \pm 0.7) \cdot 10^{-3}$ of all the detections can be explained by blazars in our sample. This is around 1/3 of the total amount of astrophysical neutrinos in the three-year dataset \citep{1908.09551}, similar to the fraction for $E_\nu\geq200$~TeV case (\citealt{neutradio1} and \autoref{s:joint_analysis} here).
Additionally we explore the dependence of the excess neutrino count on their energy. This count continues growing until the lowest energies available in the three-year IceCube dataset, on the order of 1~TeV, suggesting that many blazar-associated neutrino detections are in the range of 1-10~TeV. However, the statistical errors in the estimates based on three years of observations are too large to reliably conclude whether this is the case and to give more precise estimates. These lower-energy events contribute little to the likelihood $L$ published for the seven-year data set.

\subsection{Neutrino Energy Flux: Seven Years of Data}
\label{s:flux:energy}

Here, we present an order-of-magnitude estimate of the total neutrino flux that can be explained by radio-bright blazars. Our observational results in \autoref{s:analysis_logp} based on the IceCube $L$ map indicate that at least 70 blazars above 0.33 Jy are associated with neutrinos detected over seven years. This yields a lower limit of 10 detected neutrinos per year coming from the ensemble of bright blazars with $S > 0.33$~Jy, even if each source only resulted in a single neutrino detection. The $L$ map represents the likelihood of an astrophysical-neutrino point source and should be dominated by the contribution from events above 40~TeV \citep{1908.09551}. Lower-energy neutrinos associated with blazars also exist (see \autoref{s:flux:count}), but they get strongly downweighted in the computation of the map. We take the energy of 40~TeV as a reasonable estimate for a typical event affecting significantly the value of $L$. The effective area of IceCube at such energies is about 30~m$^2$  \citep{icecubecollaborationAllskySearchTimeintegrated2017}. Under these assumptions, we obtain a lower limit on the muon neutrino flux from these objects over the entire sky: $F_\nu^{\rm > 0.33\,Jy} \gtrsim 80 \ \text{eV}\ \text{cm}^{-2}\ \text{s}^{-1}$. According to \citet{1908.09551}, the total astrophysical muon neutrino flux at energies above 40~TeV is $F_\nu^{\rm total} \approx 835 \ \text{eV}\ \text{cm}^{-2}\ \text{s}^{-1}$. Thus our crudely estimated $F_\nu^{\rm > 0.33\,Jy}$ already constitutes almost $10\%$ of $F_\nu^{\rm total}$. Account of other blazars from our sample, with lower flux densities down to 0.15~Jy, would result in a value $F_\nu^{\rm > 0.15\,Jy} \approx (1.5\dots2.5) \cdot F_\nu^{\rm > 0.33\,Jy}$. The coefficient depends on the specific relationship, or lack thereof, between the radio and neutrino fluxes of the blazars. These estimates show that $F_\nu^{\rm > 0.15\,Jy}$ makes up for 1/4 of $F_\nu^{\rm total}$. If, in fact, the 70 associated blazars emitted multiple neutrinos on average, then this fraction gets proportionally increased. Moreover, some blazars within our sample might not get associated with neutrinos because they, e.g., fall into an area with a higher IceCube background. Our estimates are effectively lower limits, and the real number of neutrino-blazar associations in the sample may be larger.

The estimates in this subsection and in \autoref{s:flux:count} are based on very different approaches and observational datasets from IceCube. Yet the results turn out to be similar and qualitatively consistent. Remaining differences, if any, can be explained by different energies effectively contributing to each dataset.
Both of the analyses imply that it is possible to explain the entire astrophysical neutrino flux, as it is estimated from muon-track IceCube studies \citep{IceCube-1607.08006,1908.09551}, by blazars hosting radio-bright parsec-scale jets, without requiring any extreme assumptions.
Indeed, many similar radio blazars are fainter than the flux density limit of our sample, either because of their intrinsic power, or due to geometrical properties of the beamed emission, but in many cases simply because they are more distant. Together they are expected to contribute a major fraction to the total neutrino flux, though the limitations of the present analysis do not allow deriving it precisely.

\subsection{Neutrino Luminosity of a Blazar}
\label{s:flux:luminosity}

Next, we estimate the total power of neutrinos emitted by a typical blazar among those associated with IceCube detections in \autoref{s:analysis_logp}. For blazars brighter than 0.33~Jy, the realistic situation is somewhere in between the two possible opposite scenarios. The first possibility is that only 70 of such blazars emit neutrinos: in this case, we found all of them in \autoref{s:analysis_logp}. The second possibility is that all 700 bright blazars emit neutrinos at similar rates, and the 70 associated ones just happened to be detected in the covered time period.
Corresponding estimates on the average per-source neutrino flux range from 1/70 to 1/7 neutrinos per year if a single neutrino from each of those blazars got detected. Adopting the same assumptions regarding neutrino energies and the IceCube effective area as in \autoref{s:flux:energy}, we obtain the observed neutrino flux from an individual blazar $F_\nu\approx (0.06\dots0.6)\, \text{eV}\, \text{cm}^{-2}\,\text{s}^{-1}$.

Typical VLBI-selected blazars are located at $z \sim 1$ \citep{2019ApJ...874...43L}.
The median jet opening angle is estimated by \citet{JetAngles} as $1.3^\circ$, and the median relativistic beaming angle is about $5^\circ$ \citep[see Lorentz factor estimates in][]{2019ApJ...874...43L}.
We assume that neutrinos, like photons, are emitted isotropically in the emission region frame. If the region is associated with the jet, then
neutrinos are emitted within the angle of about $6^\circ$ from the jet direction. 
We derive that the total all-flavor neutrino luminosity of such a blazar is approximately $L_\nu\approx(4\cdot10^{42}\dots4\cdot10^{43}) \, \text{erg}\,\text{s}^{-1}$. Both estimates are a few orders of magnitude below the typical bolometric luminosity of bright blazars, $L_{\rm bol} \sim 10^{45}\ \text{erg} \ \text{s}^{-1}$ \citep[e.g.,][]{2002ApJ...579..530W}. These neutrino luminosities, low compared to $L_{\rm bol}$, are expected for radio quasars in order to satisfy constraints from the lack of observation of individual bright neutrino sources \citep{MurWaxMultiplets,NeronovEvolution}.

\section{Physical implications} \label{s:discussion}

Together with \citet{neutradio1}, the present study ties the neutrino production to central parsec-scale regions of radio-bright blazars, now with an even higher significance and for a wider range of neutrino energies. This gives more constraints on the neutrino production mechanism. Generally, high-energy neutrinos may be produced in hadronic (proton-proton, $pp$) or photohadronic (proton-photon, $p\gamma$) interactions.
In bright central parsecs of blazars, $pp$ interactions are suppressed with respect to $p\gamma$ \citep{Sikora1987}, though the degree of the suppression depends on the particle energy \citep[e.g.,][]{inoue,MurMeV}.

\begin{figure*}
    \centering
    \includegraphics[width=0.8\linewidth]{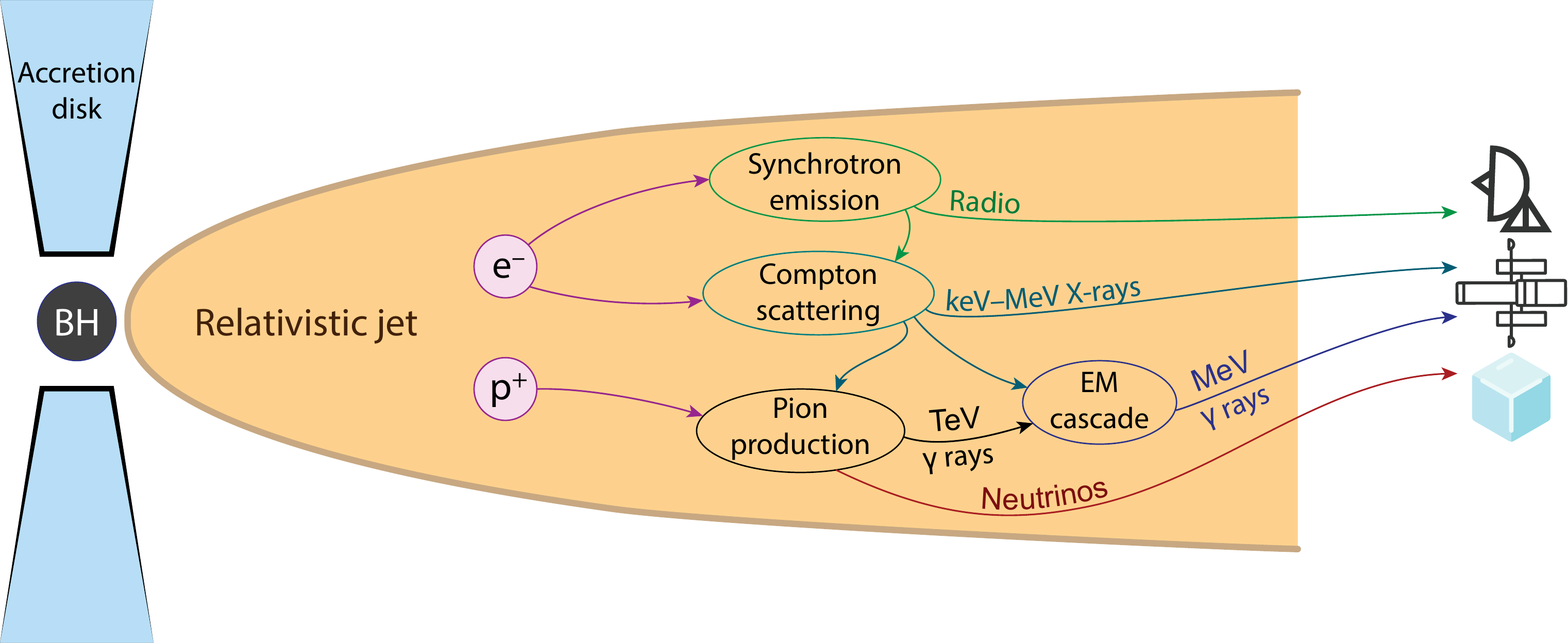}
    \caption{Production of neutrinos and relevant electromagnetic (EM) emission in blazars' relativistic jets at parsec scales. See motivation, discussion and numerical estimates in \autoref{s:discussion:model}.}
    \label{fig:production_diagram}
\end{figure*}

Here, we briefly summarize some implications of our results in a phenomenological context. These are the keys to understanding the actual mechanism of the neutrino production. \\
1. Association of neutrino arrival directions with VLBI-selected blazars suggests the importance of Doppler-boosted jets in the neutrino sources.\\
2. Neutrino events have been shown to correlate with month-scale radio flares. This motivates the assumption that neutrinos may be produced in the observed jet base. \\
3. Neutrinos with observed energies from TeVs to several PeVs most likely originate in objects of one class, blazars. Within the $p\gamma$ scenario, this requires the presence of target photons whose energy range is determined below.\\
4. Previous studies did not reveal any significant statistical association of neutrinos with GeV gamma-ray emission of blazars, either in a steady state or during flares. This may be explained if the bulk of the GeV photons are emitted in different regions than neutrinos. Target photons for $p\gamma$ interactions may also be not connected with the observed gamma-ray emission.\\
5. Individual blazars do not reveal themselves as significant neutrino sources in the present data. At the same time, our statistical analysis of the entire sample makes it possible to establish the association and demonstrates that a large number of blazars contribute to the neutrino flux, consistent with the origin of the entire astrophysical neutrino flux in sources of this class.

Though much more studies are required to construct a working quantitative model of the neutrino production, we present a possible scenario that qualitatively explains the observational data 
and emphasizes the connection between the radio emission and neutrino production.

\subsection{Possible Neutrino Production Mechanism}
\label{s:discussion:model}

\subsubsection{Self-Compton Target Photons}
\label{s:discussion:model:SSC}

In the first approximation, the cross section of the $p\gamma$ process is saturated by the $\Delta$ resonance that constrains the product of the energies of the proton and the photon in the emission region frame, while the proton energy is always $\sim20$ times that of the neutrino \citep[e.g.,][]{Derm-Menon-book}. Taking the Doppler boosting into account, 
one finds that the required emission-frame energies of the target photons are $\sim200$~eV to 200~keV for 1~TeV to 1~PeV observed neutrinos. These target-photon energies are far too high for the external photons from the accretion disk, which were previously invoked to explain neutrinos of PeV energies from radio quasars \citep{Kalashev:2014vya,DermerMuraseInoue}. A possible exception is the tail of the corona emission \citep{inoue,MurMeV}. We have only studied neutrinos above 200 TeV in \cite{neutradio1} and could not make such conclusions there due to a narrower range of required photon energies.

The observed radio emission from blazars is the synchrotron radiation from a population of non-thermal electrons in the jet. It is inevitably accompanied \citep[e.g.,][]{Derm-Menon-book} by the synchrotron self-Compton (SC) radiation, consisting of the synchrotron photons upscattered to high energies by the same relativistic electrons. The SC photons may or may not dominate the observed flux at high energies, but they must always be present. We propose a scenario where they play the role of target photons for neutrino production. The typical energies of SC photons lie in the keV--MeV range, matching the $\Delta$-resonance requirements. Moreover, the observed temporal correlation between blazar radio flares and neutrino arrival \citep{neutradio1,hovatta2020} suggests a physical connection between synchrotron and neutrino emission processes. SC radiation providing target photons for the p$\gamma$ process can be this connection. The overall mechanism of neutrino and photon production, including Compton scattering, is schematically illustrated in \autoref{fig:production_diagram}.

Note that the production of observed GeV--TeV radiation in gamma-ray bright blazars is typically unrelated to the SC processes. Emission at these energies is associated with even more compact regions closer to the central black hole, as determined, e.g., from the day-scale variability \citep{2006Sci...314.1424A,2015ApJ...807...79H}. There, inverse-Compton scattering of external photons may be important. These photons come from outside the jet and originate in the accretion disk, its hot corona, broad-line region and dust torus. This emission becomes less important at $\sim 10$~pc from the black hole, where the radio-emitting blobs are typically observed \citep{2010ApJ...722L...7P,2012A&A...545A.113P}. In addition, the external radiation is redshifted in the jet frame, which makes those photons hardly relevant for the neutrino production there.

We provide basic numerical estimates demonstrating that the proposed mechanism is feasible for neutrino production in a typical bright blazar. First we calculate the SC photons density based on the observed X-ray flux, and then utilize this density to infer constraints on protons in the following subsection.
The differential density $n_\gamma'(E'_\gamma)$ of target photons with energies $E'_\gamma$ in the emission-region frame is related \citep{Derm-Menon-book} to the observed photon flux at the observed energy $E_\gamma$,
 \begin{equation}
 \mathcal{F}_\gamma(E_\gamma)=\frac{\delta^4c}{3 d_L^2} E_\gamma'^2 r_b'^2 n_\gamma'(E'_\gamma),
 \label{*}
 \end{equation}
where $\mathcal{F} \equiv E\, dF/dE$ is the flux density measured at the Earth,

$\delta$ is the jet Doppler factor, $d_L$ is the source luminosity distance and $r_b'$ is the emitting region radius in its frame of reference. We utilize typical values of $r_b'\approx 0.2$~pc, the source redshift $z\approx 1$ and $\delta\sim 10$ \citep{2019ApJ...874...43L}. The required target-photon energy to produce neutrinos with the observed energies $E_\nu \approx 40$~TeV is then $E'_\gamma \approx 10$~keV, corresponding to $E_\gamma \approx 48$~keV. 
We estimate the typical observed flux $\mathcal{F}_\gamma(E_\gamma)$ at this energy by relating it to the radio flux through average broadband spectra of blazars \citep{Ghiss1,Ghiss2,Padov}: they imply $\mathcal{F}_\gamma(50~\mbox{keV}) \simeq (20 \dots 200) \mathcal{F}_\gamma(8~\mbox{GHz})$. For typical radio blazars from our sample this results in $\mathcal{F}_\gamma(E_\gamma)\sim (0.2 \dots 2) \cdot 10^{-11}$~erg\,cm$^{-2}$\,s$^{-1}$. Note that this value matches well the observations for individual sources, such as 3C~279 \citep{mwl3C279}, which was associated with a neutrino event by \citet{neutradio1}. Based on an average value of $\mathcal{F}_\gamma(E_\gamma)\sim 8 \cdot 10^{-12}$~erg\,cm$^{-2}$\,s$^{-1}$,
one obtains the number density of target photons $E'_\gamma n'_\gamma(E'_\gamma)\sim 2\cdot 10^4~\mbox{cm}^{-3}$.

\subsubsection{Accelerated Protons}
\label{s:discussion:model:protons}

The other ingredient of the $p\gamma$ process is the population of non-thermal protons. For observed neutrino energies $E_\nu$ between 1~TeV and 1~PeV and assuming the $\Delta$-resonance production channel, the required proton energies in the emitting-region frame $E'_p\approx 20E'_\nu (1+z)/\delta$ are in the $\sim$10~TeV to 10~PeV range. These energies are easily reached in shocks close to the base of the radio jet if these shocks are mildly relativistic \citep{Bykov-accel-1205.2208,Waxman-0907.1354}. Such slow shocks are expected to be present in the jet launching region \citep[see discussion of stationary and moving shocks in][]{1988ApJ...334..539D,1997ApJ...482L..33G,2001ApJ...549L.183A,2006MNRAS.367..851C,0901.2578,2009ApJ...696.1142M,2016A&A...588A.101F} and are often observed by VLBI close to the jet base as stationary or slow features \citep[e.g.,][]{2012ApJ...758...84P,2013A&A...551A..32F,2013ApJ...772...14C,2017ApJ...846...98J,2019ApJ...874...43L,2020MNRAS.495.3576K}. 

The proton-photon interactions are dominated by the $\Delta$-resonance contribution with the cross section of $\sigma\approx 500~\mu$b. Thus the typical density of comptonised synchrotron photons obtained in \autoref{s:discussion:model:SSC} implies the proton interaction rate $\sigma n \sim 3 \cdot 10^{-5}\, \mbox{pc}^{-1}$. This allows us to estimate the injected power of protons in an individual source as $L_p\sim 20 L_\nu/(l\sigma n),$ where $L_\nu$ is the neutrino luminosity, and $l$ is the path length of the proton. For stationary or slow shocks at the jet base, $l$ is determined by the shock lifetime. We assume $l$ is at least 100~parsecs or $300$~light years because of decades of observations of bright blazar cores and nearby stationary knots with VLBI without signs of decline.  
Given the estimated neutrino luminosity of contributing sources (\autoref{s:flux:luminosity}), we obtain $L_p\sim 3 \cdot 10^{46}$~erg/s.
This value exceeds the typical observed bolometric luminosity of sources in our sample, cf.\ Section~\ref{s:flux:luminosity}, but is below the Eddington luminosity for a black hole mass of $\sim 10^9 M_\odot$ typical for radio blazars \citep{2012A&A...545A.113P}. Thus the injected power of protons does not need to exceed the Eddington power by a large factor, unlike in proton-jet models \citep{Boettcher-p-blazar}.

We note that the required proton energies can also be reached by direct acceleration in the black-hole magnetosphere: see the ``high-luminosity'' regime in \citet{magnetosphere-Ptitsyna}, and, e.g., \citet{2020PhRvD.102d3010I} for other approaches. Transfer of these protons to the jet would, however, require a special mechanism \citep{Neronov-Aharonian}. 

\subsubsection{Resulting Neutrino Spectrum}
\label{s:discussion:model:nu-spectrum}

The spectral shape and the total neutrino flux from a blazar are directly related to those of the target photons, so the neutrino spectrum from an individual blazar is not power-law even if the proton spectrum is \citep[see, e.g.,][]{Stecker:1991vm,Kalashev:2014vya}. Different blazars have SC bumps at different energies \citep[e.g.,][]{2010ApJ...716...30A,2016ApJS..224...26M}, and the study of a complicated blazar population is required to predict the overall observed neutrino spectrum. In particular, BL~Lacs have higher peak frequencies than radio quasars and are therefore more important for the production of neutrinos of lower energies. Note that extreme BL Lacs, also proposed earlier as the source of high-energy neutrinos \citep{extremeBL}, may have the synchrotron SED peak at the required target-photon energies, but they are much less numerous than the blazars we study here. 

\subsection{Lack of Gamma-Ray Associations} 
\label{s:discussion:model:no-gamma}

Together with neutrinos, gamma rays of similar energies are born in the same $p\gamma$ interactions. They, however,
can produce electron-positron pairs on 
background photons, giving a start to the electromagnetic cascade: see \autoref{fig:production_diagram} for the process schematics. 
Two photons with energies $E'_1$ and $E'_2$ in the emission-region frame are able to produce a $e^- e^+$ pair provided their center-of-mass energy exceeds the rest energy of the pair: $E'_1 E'_2 \ge 2 m_e^2c^4$. The pair-production cross section $\sigma_{\gamma\gamma}$ peaks strongly just above this threshold and the process is saturated \citep{Derm-Menon-book} by scattering of photons with $E'_1 E'_2 \approx 4 m_e^2c^4$, for which $\sigma_{\gamma\gamma}\simeq  2 \cdot 10^{-25}~\mbox{cm}^2$.

Blazar jets contain enough background photons for this cascading to be efficient at energies starting from a fraction of a GeV. To show this, consider photons with observed energies $E_1\sim 0.5$~GeV, close to the lower-energy threshold of the Fermi LAT. For them, the relevant energy $E'_2\approx \frac{4m_e^2c^4}{E_1}\, \frac{\delta}{1+z}\approx 10$~keV, similar to the energy of target photons for the $p\gamma$ process described in Section~\ref{s:discussion:model:SSC}. The mean free path of a 0.5-GeV photon in the jet frame is therefore $l'_\gamma=1/(\sigma_{\gamma\gamma}n'_\gamma(E'_\gamma))\approx 70$~pc. Higher-energy photons produce pairs on background photons with lower energies, $E'_2 < 10$~keV, which are more numerous: $n'_\gamma$ is larger, cf.\ Eq.~(\ref{*}). Since the cross section $\sigma_{\gamma\gamma}$ stays the same, the mean free path is even shorter for photons of higher energies.

Emitted photons travel within the jet until leaving it towards the observer. This path length can be estimated geometrically as the jet depth along the line of sight close to the radio-emitting core. Typical values of the jet opening angle ($1.3^\circ$, see \citealt{JetAngles}), the jet viewing angle ($2^\circ$, see \citealt{2019ApJ...874...43L}), and the distance from the radio core to the jet apex (23~pc, see \citealt{2012A&A...545A.113P}) yield the depth $l_{\rm jet}\sim 15$~pc in the observer frame. This corresponds to $l'_{\rm jet}=l_{\rm jet}\delta/(1+z) \sim 75$~pc in the jet comoving frame. The background photon density stays similar along this path, thus the optical depth with respect to pair production is $\tau \approx l'_\gamma/l'_{\rm jet} \gtrsim 1$ for gamma rays above $0.5$~GeV.

Therefore the gamma rays resulting from the electromagnetic cascades started from the $p\gamma$ interactions are hardly observable by \textit{Fermi} LAT. This lack of strong direct neutrino -- gamma-ray relation is also supported by multimessenger constraints on sources of neutrinos of these energies \citep{MurHidden}. Observed GeV gamma-rays originate closer to the black hole \citep{2010ApJ...722L...7P} and do not pass through radio-emitting regions.
In the MeV band individual measurements of the blazars' fluxes are mostly absent. Even when the interpolation between keV and GeV is possible, it is not clear which part of the flux comes from the compact jet observed by VLBI and which is born closer to the central engine.
Results of future missions aimed at the (sub)MeV astronomy, e.g., eASTROGAM \citep{eASTROGAM}, and AMEGO \citep{2019BAAS...51g.245M}, would be important for obtaining refined quantitative predictions about the neutrino emission. 
Note, however, that for many sources this contribution to the gamma-ray luminosity is subdominant. Indeed, the neutrino luminosity, and hence the cascade-photon luminosity, is much smaller than the bolometric photon luminosity of the source, often saturated by hard gamma-ray emission. Together with the strong cosmological evolution of radio blazars \citep{2009ApJ...696...24S,2017A&A...602A...6S}, this relaxes \citep{NeronovEvolution} standard constraints on neutrino-emitting quasars from non-observation of individual sources.

\section{Summary} \label{s:summary}

Central parsecs of radio-bright blazars were shown previously \citep{neutradio1} to be associated with astrophysical neutrinos above 200~TeV detected by IceCube. In this work, we analyze newly available information about muon-track events from 2008-2015 with energies from a fraction of a TeV to a few PeVs. We demonstrate that the neutrino-blazar association holds for the entire energy spectrum. The combined post-trial significance of directional correlations found in two independent analyses, at lower and higher energies, corresponds to a chance probability of $p = 4\cdot10^{-5}$ ($4.1\sigma$).

The extension of the neutrino-blazar association to lower energies changes our understanding of the neutrino production mechanism: higher-energy target photons are required to produce lower-energy neutrinos in the $p\gamma$ interactions, which is the most probable channel of neutrino production in blazars. In radio-loud blazars, these target photons may be provided by the X-ray self-Compton radiation, which inevitably accompanies the synchrotron radiation of non-thermal electrons observed in the radio band from the parsec-scale jet. High-energy neutrino emission and gamma radiation may be, to an extent, independent: they may be produced in different zones of the central parsecs in blazars. This explains the lack of association between gamma-ray loud blazars and IceCube neutrinos reported in numerous previous studies.

The association we report here was found on statistical and not on event-by-event grounds. It implies that the sources are numerous and most of them do not stand out individually in the seven-years IceCube sample.
We expect that even in future studies with larger statistics, any analysis focused only on the brightest spots in the neutrino map would miss most of the sources. Many actual neutrino sources still remain outside of our flux-limited sample, including distant or less beamed blazars. Neutrino luminosity of an individual source is orders of magnitude lower than its bolometric photon luminosity.
Overall, we explain at least 1/4 of the astrophysical muon neutrino flux, derived from IceCube track data \citep{1908.09551}, by the VLBI-selected blazars brighter than 0.15~Jy. This is consistent with the entire neutrino flux being produced in central parsecs of radio-bright blazars. However, cascade observations may indicate that the flux may be higher at dozens of TeV \citep{2020arXiv200109520I}, and this excess may be associated with a different component \citep{Vissani2comp,AhlersHalzen}.

Future studies will help to verify and clarify the relation between radio quasars and high-energy neutrinos. The results of the present work can be tested with the full collected IceCube dataset. 
Since 2020, IceCube alerts are followed by immediate radio observations by VLBA and RATAN-600. Independently, a set of probable high-energy neutrino emitters is continuously monitored by the same instruments. In the nearest future, the study will be extended to Baikal-GVD \citep{BaikalGVD} neutrino candidate events. Further ahead, KM3NeT \citep{KM3NeT} and PONE \citep{2020NatAs...4..913A} in neutrinos, eASTROGAM \citep{eASTROGAM}, AMEGO \citep{2019BAAS...51g.245M}, and SRG \citep{2011SPIE.8147E..06P} in keV to GeV photons will supplement these studies with important multimessenger information.

\acknowledgments

We thank the anonymous referee, Markus B\"ottcher, Andrei Bykov, Timur Dzhatdoev, Marcello Giroletti, Matthias Kadler, Alan Marscher, Kohta Murase, Elena Nokhrina, Manel Perucho, Egor Podlesny, Eduardo Ros, Bair Shaibonov for helpful comments and discussions. 
We are grateful to Elena Bazanova for English language editing and proofreading of the text.
This work is supported in the framework of the State project ``Science'' by the Ministry of Science and Higher Education of the Russian Federation under the contract 075-15-2020-778.
This research has made use of the NASA's Astrophysics Data System.
\facilities{IceCube neutrino observatory, VLBA, EVN, LBA.}

\appendix

\section{IceCube all-sky likelihood map} \label{app:icecube_map}

Our statistical analysis is based on the IceCube sky map of pre-trial local $p$-values that represent the neutrino point-source likelihood for each direction (\autoref{s:data_icecube}). In this paper we denote the logarithms of those $p$-values as $L = -\log p$. We have noticed certain features of the value distribution in this map, and briefly discuss them in this appendix. Additionally, we attempt to minimize an effect they might have on the analysis.

\begin{figure}
    \centering
    \gridline{\fig{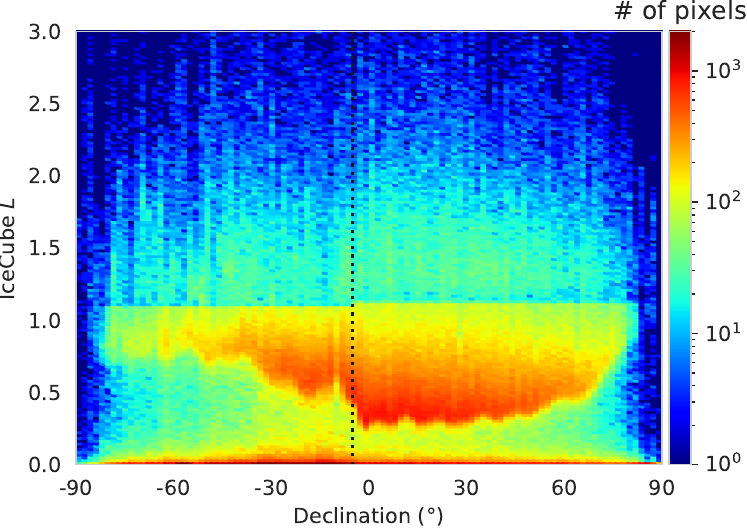}{0.5\textwidth}{\raggedright(a) Two-dimensional histogram of IceCube $L$ distribution for each declination in the sky. The color indicates the number of pixels at each declination having the corresponding values of $L$. The total number of pixels is 3145728; less than 0.1\,\% of them have $L > 3$ and are not shown in this histogram. The region to the right of the vertical dashed line, $\delta > -5^\circ$, is used in our analysis.}
              \fig{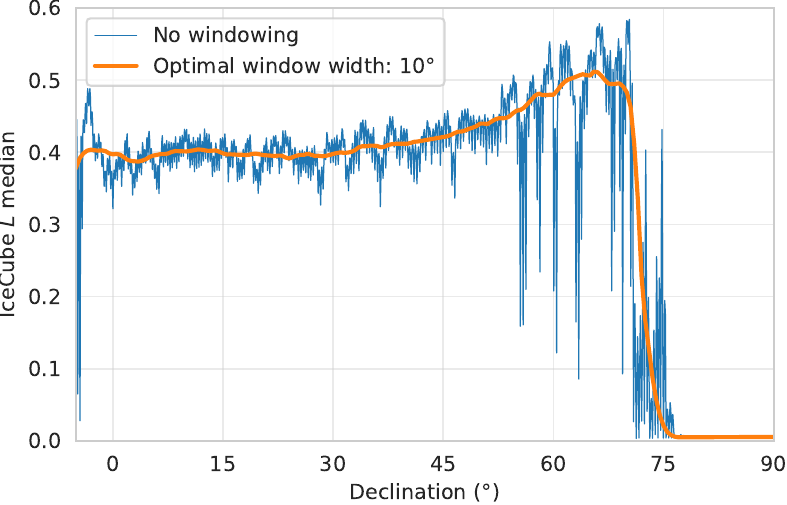}{0.5\textwidth}{\raggedright(b) Median filtering of the $L$ values with respect to the declination. The blue line marks the median for each individual declination, i.e., each row of pixels in the map, separately. The orange line corresponds to the median computed with the optimal window width of $10^\circ$ around each declination value.}}
    \caption{Illustration of the $L$ distribution in the IceCube seven year sky map.}
    \label{fig:icecube_map}
\end{figure}

The histogram of $L$ values across the full range of declinations is shown in the left panel of \autoref{fig:icecube_map}. A prominent feature in this histogram is the abrupt jump around $L \approx 1$: the values of $L$ just below the cutoff occur about an order of magnitude more often than $L$ values immediately above. This jump is present for all declinations outside of narrow polar regions, and the cutoff value differs between the northern ($\delta > -5^\circ$) and the southern sky ($\delta < -5^\circ$). We acknowledge that such a peculiarity is present in the $L$ value distribution, but believe that it does not affect our study in any meaningful way. We only utilize the Northern sky, as justified in \autoref{s:data_icecube}, and the jump of the $L$ distribution stays constant in this declination range.

The right panel of \autoref{fig:icecube_map} presents the median value of $L$ for each declination. As apparent from this plot, the typical values of $L$ are highest at declinations around $60^\circ$-$70^\circ$. If taken at a face value, this effect would indicate that more astrophysical sources stand out from the background at these high declinations. However, the effective sensitivity of IceCube is the highest within $\delta \in [-5, 45]^\circ$ \citep{icecubecollaborationAllskySearchTimeintegrated2017}. The increased $L$ values further to the north should be treated with care. To reduce their effect on any further analysis as conservatively as possible, we first perform a median filtering (running median) on the $L$ values with respect to declination. The filter's window width is chosen from values between $0^\circ$ and $30^\circ$ in steps of $1^\circ$ by minimizing the mean absolute error via a cross-validation approach; the optimal value turned out to be $10^\circ$. The filtered $L_{\rm med}$ are shown in the right panel of \autoref{fig:icecube_map} as well. Then we define the \textit{normalized $L$ values:} $L_{\rm norm} = L - L_{\rm med}$, where $L_{\rm med}$ corresponding to the pixel's declination is subtracted from the original value in each pixel. Only these normalized values $L_{\rm norm}$ are used in other sections of this paper. Our approach to account for the effect of higher $L$ values at higher declinations increases the robustness of the results of the performed statistical analysis.

As can be seen in both panels of \autoref{fig:icecube_map} and in the map itself (\autoref{fig:skymap}), the typical $L$ values drop closer to zero when approaching the poles; in the Northern sky this change happens around $\delta = 75^\circ$. Nevertheless, we do not treat this area specially in any way and believe that it does not have any significant effect on any further analysis: only $\approx 3\%$ of the Northern sky is in the region $\delta > 75^\circ$.

\bibliography{neutradio}
\bibliographystyle{aasjournal}

\end{document}